# Change-over of carrier type and magneto-transport property in Cu doped $Bi_2Te_3$ Topological Insulators

Abhishek Singh[1], Rahul Singh[1], A. Lakhani[2], T. Patel[2], G. S. Okram[2], V. Ganeshan[2], A. K. Ghosh[3] and Sandip Chatterjee[1,*]

[1]Department of Physics, Indian Institute of Technology (Banaras Hindu University), Varanasi-221005, India

[2]UGC-DAE Consortium for Scientific Research, Indore, Madhya Pradesh 452001, India

[3]Department of Physics, Banaras Hindu University, Varanasi-221005, India

## Abstract

Structural, resistivity, thermoelectric power and magneto-transport properties of Cu doped $Bi_2Te_3$ topological insulators have been investigated. The occurrence of the tuning of charge carriers from n type to p type by Cu doping at Te sites of $Bi_2Te_3$ is observed both from Hall effect and thermoelectric power measurements. Carrier mobility decreases with the doping of Cu which provides evidence of the movement of Fermi level from bulk conduction band to the bulk valence band. Thermoelectric power also increaseswith doping of Cu.Moreover linear magnetoresistance (LMR) has been observed at high magnetic field in pure $Bi_2Te_3$ which is associated to the gapless topological surface states protected by time reversal symmetry (TRS), whereas doping of Cu breaks TRS and an opening of band gap occurs which quenches the LMR.

Topological insulators (TIs) are a new class of materials, which are insulating in bulk but conducting at thesurface. This is due to the gapless edge or spin resolved surface states (SS), which are topologically protected by time reversal symmetry. The spins are locked in the perpendicular direction of momentum due to the strong spin-orbit interaction. As a matter of fact, electrical conduction is robust against backscattering at the edge states or on the surfaces in TIs. These special helical spin properties of electrons make TIs interesting and relevant for new physics. Since the locking of spin and orbital states is protected by time reversal symmetry,[1] the delocalized surface states are unaffected from nonmagnetic dopants and defects. The possibility of Majorana Fermions,[2] topological superconductivity,[3,4]novel magnetoeleric quantum states,[5]the absence of backscattering from nonmagnetic impurities,[6]exciton condensation,[7] magnetic monopole,[8]and anomalousquantum Hall effect[9] types of exotic properties in TIs are very promising in the application of spintronic devices and quantum computing. Topological surface states in $Bi_2Te_3$ and $Bi_2Se_3$ with only one massless Dirac cone on each surface have been studied using Angle-resolved photoemission spectroscopy (ARPES).[10,11]Quantum magneto-transport phenomenon such as weak antilocalization,[12-14]Aharonov-Bohm oscillations,[15] and quantum conductance fluctuations,[16] are associated with surface states. The time reversal symmetry protection of the Dirac point can be lifted by magnetic dopants, resulting in a band gap due to the separation in the upper and lower branches of the Dirac cone.[17]

Moreover, materials with large magneto resistance (MR) have been of great interest from the application point of view as well as for fundamental research. According to Abrikosov'stheory,[18]a quantum linear magneto resistance which is ideal for application, is expected in the materials having zero band gap with linear energy spectrum. Materials with linear MR might be used in magneto electronic applications such as multifunctional electromagnetic application and disc reading heads. Recently, a giant and linear MR in TIs have been reported.[19-26] Zhang et al. showed that on ultra thin film of $Bi_2Se_3$ a surface state gap opened at the Dirac point because of tunnelling between the top and bottom surface states.[27]Interestingly, while most of the reports in $Bi_2Te_3$ samples were on doping on Bi site,[28-32]doping on Te site has not yet been reported to the best of our knowledge, which however might be equally interesting to study. Here we report a study of 5% Cu doping on Te site in $Bi_2Te_3$. It is observed from Hall resistivity and thermoelectric data that type of carrier changes from n type to p type with Cu doping. Also the magneto resistance changes from linear to non-linear behavior on doping.

Single crystals of $Bi_2Cu_xTe_{3-x}$(x=0, 0.15) were grown by modified Bridgman method. High purity powder of Bi (99.999%), Te(99.999%) and Cu (99.999%) were mixed uniformly in their stoichiometric ratios. The mixture was sealed in quartz tube after evacuating down to ~$10^{-6}$torr. The crystal growth involves cooling from $950^0$C to $550^0$C in a period of 24 hours and then annealing at this temperature for 72 hours. Silver coloured single crystals were then cooled down to room temperature slowly. The transport properties were measured by using the physical property measurement system(PPMS) from Quantum Design.Thermopower was measured using a standard home-made set up.Sample was sandwiched between two cylindrical oxygen-free highly conducting copper blocksin vacuum. The voltage difference between the cold and the hot ends was measured at each chosentemperature (details are given in Ref. 33).

Single crystals ofBi$_2$Cu$_x$Te$_{3-x}$ (x=0, 0.15)with the cleaved surface along the basal planes were characterized usingX-ray diffractionmethod collected at room temperature [Fig.1(a) and (b)]. X-ray diffracted beam directions showed only the (00L) with space group R-3m. The Full Width Half Maxima (FWHM) of the (003) peak is $0.115^0$and $0.125^0$whereas for (006) peak it is $0.131^0$ and $0.137^0$for undoped and doped samples respectivelywhich are very small indicating that as prepared samples have excellent single crystallinity in long range. Around 2θ=$54^0$ and $64^0$, a small splitting due to the difference in the wavelength of X-ray source of CuK$_α$ and CuK$_β$ radiation is observed.An additional peak around 2θ=$40^0$ for both the samples is observed which might be due to the slight misalignment or slight tilt angle of the crystallinity of as cleaved single crystals.

In order to see the electrical transport behaviour of the materials, temperature variation of resistivity under zero magnetic field for pure and Cu doped Bi$_2$Te$_3$samples were carried out[Fig.2(a)]. It is observed that the resistivity increases with temperature indicating typical metallic behaviour.Resistivity of doped sample is almost four times higher than that of the undoped sample which might be attributed to the mixture of bulk valence band and topological surface state.[34]The presence of wiggles in the resistivity data of Cu doped samples also confirms this indicting that bulk state is becoming significant above a critical temperature. Figure2(b) shows the variation of Seebeck coefficient with temperature for both the Bi$_2$Te$_3$and Bi$_2$Cu$_{0.15}$Te$_{2.85}$ samples in the temperature range of 10K-300K. Seebeck coefficient for pure $Bi_2Te_3$ sample shows negative slope whereas$Bi_2Cu_{0.15}Te_{2.85}$ sample shows positive slope in the whole range of temperature, which indicates that a change-over from*n* type conduction to*p* type conduction occurs with Cu doping in $Bi_2Te_3$. Moreover the maximum absolutevalue of Seebeck coefficient in case of $Bi_2Te_3$ is 142.8μv/K whereas in doped sample it is around 159.8μV/K. The values are consistent with those reported by Cao et.al.[35]Recently Seebeck coefficient of $Bi_2Te_3$/Cu composite has been reported.[36] It is observed as the Cu content increases the value of the Seebeck coefficient increases but it remains negative.[36] In the present investigation doping of Cu is not only tuning the carrier type but also increasing absolutevalue of Seebeck coefficient.To determine the efficiency of thermoelectric power we have calculated the power factor [shown in the inset of Fig.2(b)] of pure and undopedsamples using the formula

$$Power\ factor = S^2/\rho$$

Where $\rho$ is the electrical resistivity and S is the Seebeckcoefficient.It has already been shown that with increaseof the temperature electrical resistivityincreases in both pure and doped samples but the increase in Seebeck coefficient(S) is much larger than the increase of electrical resistivity which gives rise to the increase of power factor with increasing temperature.

In order to determine the carrier concentration (*n* or *p*) and mobility (μ) of the pure and Cu doped $Bi_2Te_3$, we have also carried out the Hall measurement at 200K and 300K. The type of carrier we have determined first and then the carrier concentration has been calculated using the slope of the Hall resistivity. Using the data of carrier concentration and electrical resistivity we determined carrier mobility of the samples.Fig.3(a) shows the variation of Hall resistivity as a function of applied magnetic fields at temperatures 200 and 300K for pure $Bi_2Te_3$. Slope of the curve is negative which shows that carriers in pure $Bi_2Te_3$ are *n* type for the entire range of temperature of measurementwhich is also consistent with the negative Seebeck coefficient of the sample.

Calculated carrier concentration (*n*) of the pure $Bi_2Te_3$ is shown in the inset1of Fig. 3(a) which comes in the range of reported value given by Zhang et.al.[37]Since topological insulators are insulating in bulk but conducting on surface, it might be that at high temperature bulk contribution is dominating over surface contribution of the sample. As we know topological surface state is a complete quantum phenomenon,existence of quantum mechanical behavior is significant at very low temperature. As a matter of fact, at very low temperature surface state is dominating over bulk state and that is why the increment in carrier concentration is very low at low temperature (T≤20K) whereas it is very high at high temperature.Moreover, the rate of increment of carrier density is also increasing with the increase of temperature, this also confirms that bulk insulating character is dominating over surface metallic character of the sample.

We have also calculated the mobility ($\mu$) of the carriers from the Hall data. Calculated mobility as function of temperature (without Field) has been shown in the inset2 of Fig.3(a).Mobility for pure $Bi_2Te_3$ sample at the applied field 2T is 271.48cm$^2$/Vsec and 131.68cm$^2$/Vsec at 200K and 300K respectively whereas at Field 5T it is 225.99cm$^2$/Vsec and 126.71cm$^2$/Vsec at 200K and 300K respectively. Hence it is clear that as we increase both the temperature and field, mobilitydecreases. This is as expected because with decreasing temperature, freezing out of phonons takes place and as a consequence, the carrier scattering and thus thermal vibration or the contribution of phonon decreases and high mobility prevails. Similar trends happen with the magnetic field.

Figure 3(b) shows the variation of Hall resistivity with the applied magnetic field for$Bi_2Cu_{0.15}Te_{2.85}$sample at 200K and 300K.The slope of the curve is positive which indicates that carriers are p type.The carrier concentration is 2.20x10$^{19}$ per cm$^3$ and 2.96x10$^{19}$ per cm$^3$ at 200K and 300K, respectively. Carrier density increases with temperature similar to the undoped sample.The variation of mobility with the applied field is shown in the inset of fig.3(b). Mobility decreases with increase in field and temperature as is observed in the $Bi_2Te_3$ and the value is lowerin $Bi_2Cu_{0.15}Te_{2.85}$ than that of the undoped sample. It may be due tothe generation of electrons by $Te_{Bi}$antisite defectsin pure sample where Fermi level lies in the conduction band. In fact, charge carriers are in conduction band and exert less resistivity in comparison to valence band and hence the mobility will be high. Doping of Cu drives the Fermi level into valence band leading to p type behavior and as the charge carriers are in valence band, more resistance(one order higher than that of pure sample)and less mobility are observed than thosein conduction band whichis also clear from the resistivity data[Fig. 2(a)]. It is worthwhile to mention that the presence of foreign element i.e. Cu may also produce extra center of scattering and resultshigher resistivity as is observed in Fig. 2. Moreover, number of charge carriers, calculated from the Hall measurement in pure sample,is higher than that ofdoped one giving rise to the low resistivity.Therefore, both Hall data and thermopowerdata corroboratethe resistivity data.

Variation in percentage of magneto resistance (MR) as a function of applied magnetic field under different temperatures for undoped and doped samples is shown in Fig. 4(a) and4(b),respectively. Fig.4 (a) shows a clear linear and non-saturating MR of $Bi_2Te_3$. MR is increasing with applied field butdecreasing with increasing temperature.As it is clear from Hall data that as we increase temperature,carrier concentration($n_0$) increasesandMR decreases. Similarly with increasing applied field(H),MRincreases [Fig.4(a)]. No saturation has been seen in pure sample whereas saturation has been observed in doped sample [Fig.4 (b)].Moreover; the %MR in undoped sample is oneorder higher than that of doped one. Li et.al. observed that when thickness of $Bi_2Te_3$ thin films is greater than 2QL(quintiplelayer) then topological surface state (TSS) appears.[38] He et al. reported that in $Bi_2Se_3$ thin films of thickness <6nm, grown by molecular beam epitaxial method, linear magnetoresistance disappears due to opening of gap in the surface state as induced by the interface coupling.[39]In $Bi_2Te_3$nanoplates having a thickness less than 5nm, absence of surface gapless state is observed by Kong et al.[40] In order to have a bulk band gap for the security of surface topological gapless state in $Bi_2Te_3$platelike samples, thickness should be greater than 14QL. Both the doped and undopedsamples have been cleaved with few mm thickness,therefore, thickness will be definitely greater than 14QL as the thickness of 1QL in $Bi_2Te_3$ is of the order of some nm, as a matter of fact thickness cannot be the reason for the absence of LMR in doped sample. Zhang et.al.[27] reported that in ultrathin film of $Bi_2Se_3$ absence of LMR has been observed because of opening of surface state gap at the Dirac point due to the tunneling between top to bottom surface states.Tuning of band gap and changing in Hall resistivity has been observedby Lee[41] in silver chalcogenides by using hydrostatic pressure. Hence, presence of LMR is possible when pressure induces closing of band gap. This shows that very small or zero band gap is essential condition for the presence of LMR. As a matter of fact the LMR is due to the presence of surface state as already reported in bulk$Bi_2Te_3$.[42]As Cu produces internal magnetic impurity i.e. magnetic field, therefore, doping of Cu changes the gapless energy spectrum into a gap. Hence, energy spectrum does not remain linear so that disappearance of LMR occurs and saturation of MR can be seen.

In conclusion, we have investigated thestructural, transport and magnetotransport properties of Cu doped and undoped $Bi_2Te_3$ topological insulators. With Cu doping, resistivity increases as Fermi level is shifted into valence bandwith extra scattering centers. It is also observed that Cu doping tunes the carrier from *n*to *p* type, which is attributed to the presence of $Te_{Bi}$ and $Bi_{Te}$ antisites. The observed LMR in $Bi_2Te_3$ single crystal is believed to be of quantum origin and associated with gapless linear energy spectrum of surface Dirac Fermions. The absence of linear magneto resistance in Cu doped sample reveals the gap opening due to the presence of magnetic impurity.

**Acknowledgement**

The authors are grateful to UGC-DAE Consortium for Scientific Research, Indore, India for providing facility. Authors are also grateful to CIFC, IIT(BHU) for providing facility for magnetic measurement. Authors also acknowledge Dr. Devendra Kumar for his help in the measurement.

**Figure Captions:**

Fig1. Room temperature X-ray diffraction patterns of (a) $Bi_2Te_3$ and (b) $Bi_2Cu_{0.15}Te_{2.85}$ single crystals.

Fig2. (a) Temperature dependence of electrical resistivity for $Bi_2Te_3$ and $Bi_2Cu_{0.15}Te_{2.85}$ single crystals showing metallic behavior and (b) Variation of Seebeck coefficient as a function of temperature. Inset represents the power factor (PF) of the $Bi_2Te_3$ and $Bi_2Cu_{0.15}Te_{2.85}$ single crystals.

Fig3. (a) Magnetic field dependence of the Hall resistivity of the pure $Bi_2Te_3$ single crystal at 200K and 300K. Inset1 represents the variation of 3D carrier concentration as a function of temperature whereas Inset 2 shows the variation of carrier mobility with temperature for the undoped $Bi_2Te_3$ and (b) Magnetic field dependence of the Hall resistivity of the $Bi_2Cu_{0.15}Te_{2.85}$ single crystal at 200K and 300K, the inset in fig is the carrier mobility as a function of applied magnetic field at 200K and 300K.

Fig4. (a) Normalized magnetoresistance $\frac{R(H)-R(0)}{R(0)}$ as a function of magnetic field for $Bi_2Te_3$ at different temperatures showing linear behavior and (b) Normalized magnetoresistance $\frac{R(H)-R(0)}{R(0)}$ as a function of magnetic field for $Bi_2Cu_{0.15}Te_{2.85}$ at different temperatures showing saturating behavior.

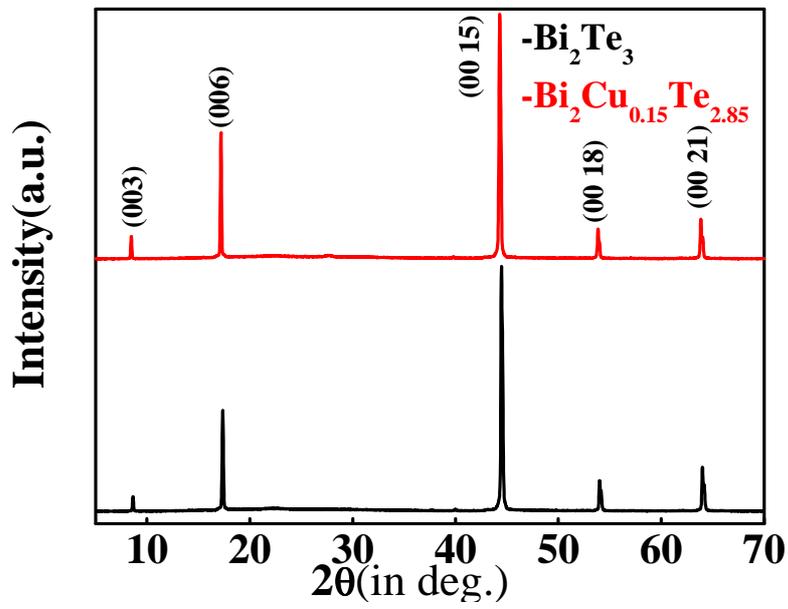

**Fig1**

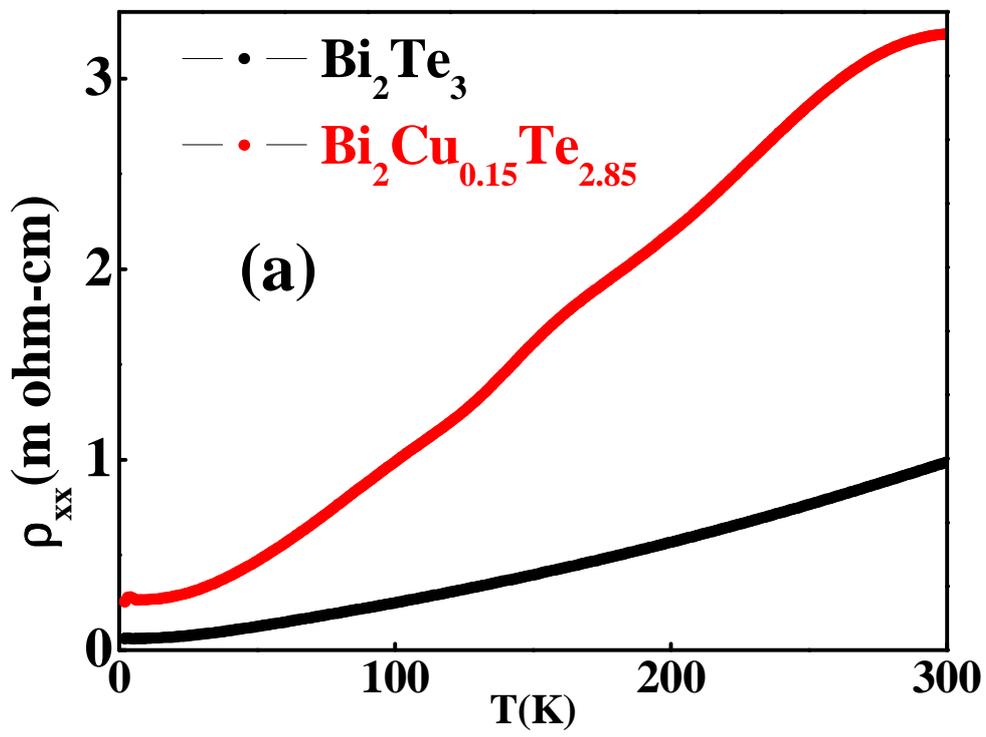

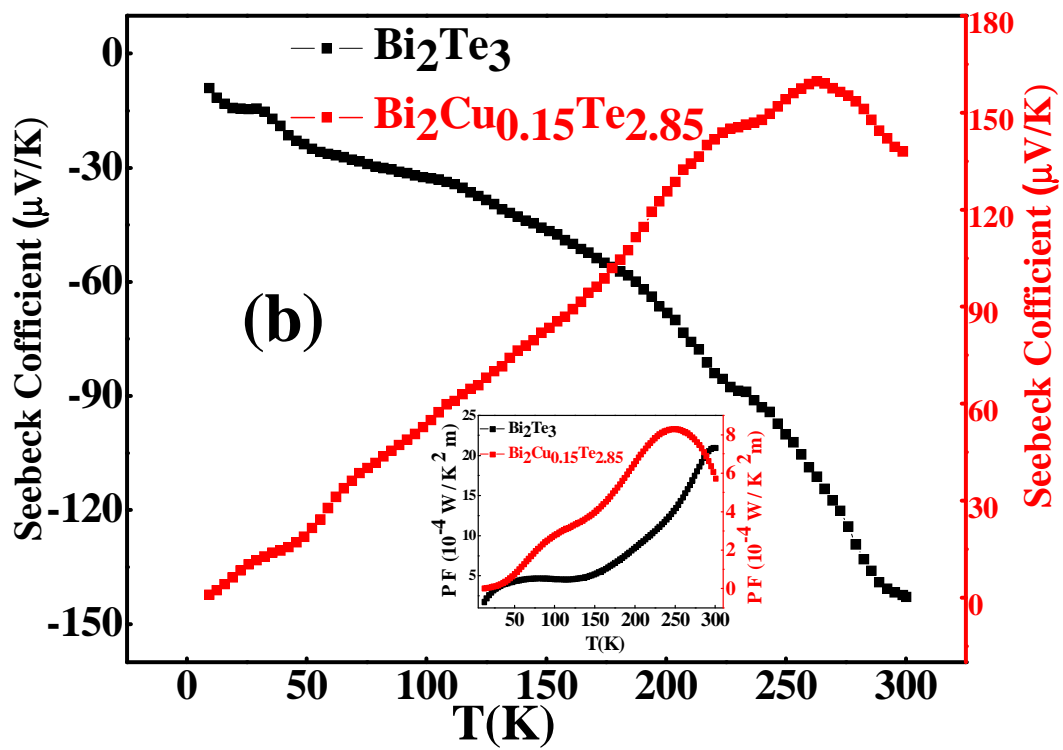

**Fig2**

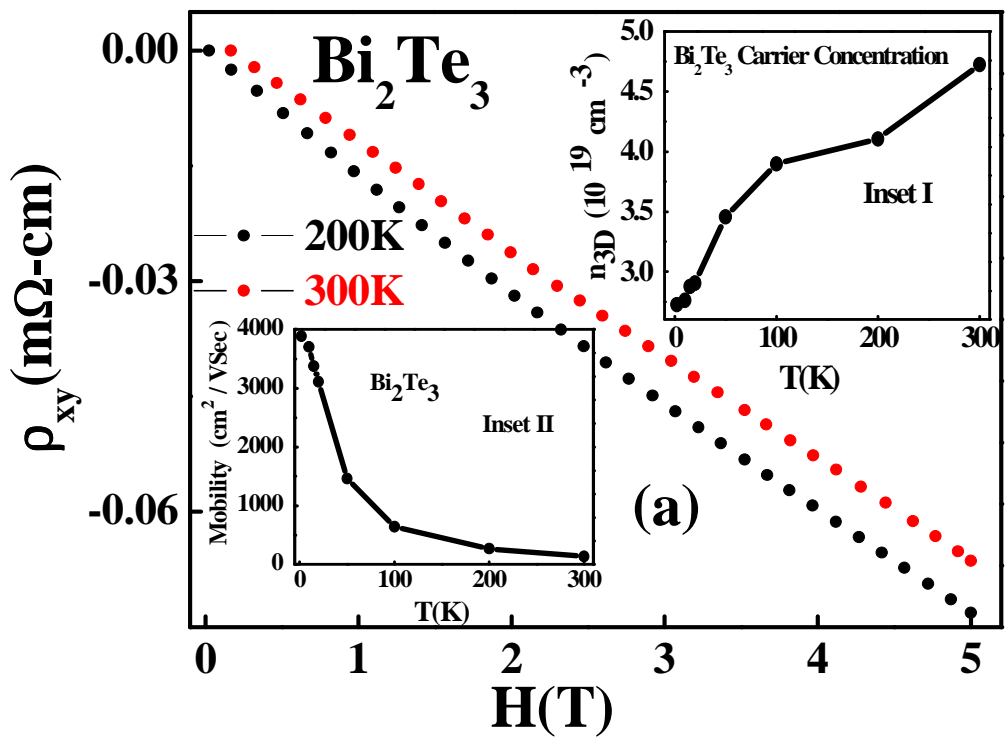

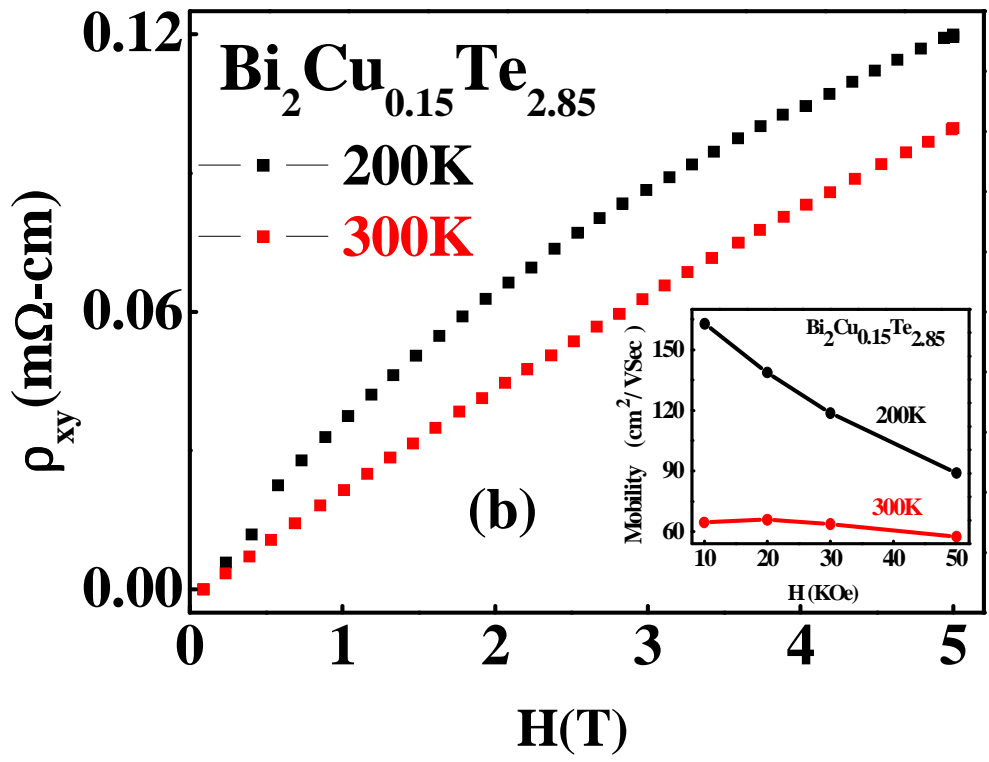

Fig3

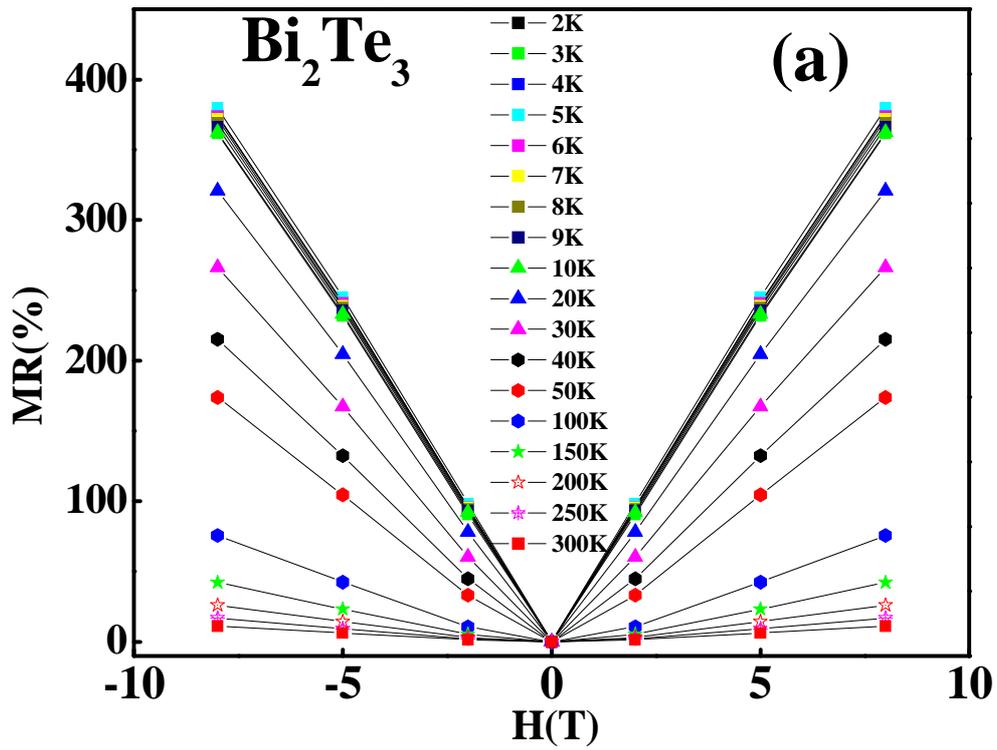

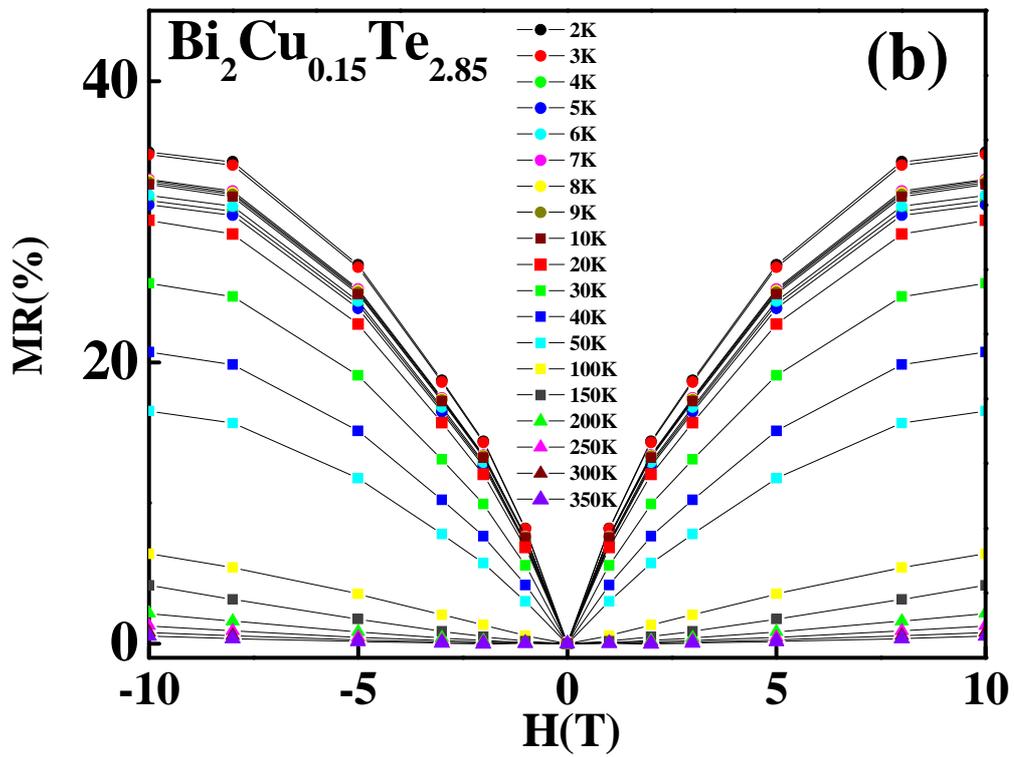

**Fig4**